\documentclass[rnote]{aa} 
\usepackage{graphicx}
\usepackage{txfonts}
\usepackage{natbib}
\bibpunct{(}{)}{;}{a}{}{,} 
\def\hda{HD~152219}

\def\heb{He\,{\sc ii}}
\def\hea{He\,{\sc i}}
\def\hbeta{H$\beta$}
\def\fer{{\sc FEROS}}
\def\mid{{\sc MIDAS}}
\def\l{$\lambda$}
\def\ll{$\lambda\lambda$}

\def\kms{km\,s$^{-1}$}

\def\degr{$^\circ$}

\defcitealias{SGR06_219}{Paper~I}
\def\papI{\citetalias{SGR06_219}}

\begin{document}
   \title{The nature of the line profile variability in the spectrum of the massive binary HD 152219\thanks{Based on observations collected at the European Southern Observatory (La Silla, Chile) under program ID 077.D-0321(A). The full version of Table 2 is available from the CDS: http://cdsweb.u-strasbg.fr}}

   \author{H. Sana
          }

   \institute{European Southern Observatory, Alonso de Cordova 3107, Vitacura, Santiago 19, Chile\\
              \email{hsana@eso.org}
             }

   \date{Received May 27, 2008; accepted March 21, 2009}

 
  \abstract
   {HD 152219 is a massive binary system with O9.5 III + B1-2 V/III components and a short orbital period of 4.2~d. Its primary component further displays clear line profile variability (LPV). The primary component being located within the pulsational instability domain predicted for high-luminosity stars, we previously suggested that the observed LPV could be associated with non-radial pulsations.}
   { The aim of the present work is to determine the nature of the observed LPV in the spectrum of the primary component of HD~152219.} 
   {During a 4-night FEROS monitoring campaign, we collected a new set of 134 high signal-to-noise spectra. These new observations were then used to re-investigate the variability of different line profiles in the spectrum of HD~152219.}
   {Based on the present analysis, we discard the  non-radial pulsations and point out the Rossiter-McLaughlin effect as the cause of the LPV in HD~152219. The upper limit on the amplitude of possible weak pulsations is set at a few parts per thousand of the continuum level. } 

    {}


   \maketitle
%

\section{Introduction}

With only a few objects known to present pulsations, asteroseismology of massive stars remains essentially a {\it terra incognita}. This partly results from the lack of dedicated observational studies and the difficulty of disentangling possible pulsations from wind effects or co-rotating structures. Among the known O-type pulsators, \object{$\zeta$ Oph} (O9~V) is the best studied object, with 19 frequencies identified either photometrically or spectroscopically and corresponding  both to radial and non-radial pulsations \citep[][and reference therein]{WKM05}. \object{HD 93521} is an isolated O9.5~Vp star exhibiting non-radial pulsations (NRPs) with periods of 1.8h and 2.9h, corresponding to harmonic degrees $l$ of 9 and 5 respectively \citep{RDBvW08, HTC98}. 
Other candidates that could harbor NRPs are \object{$\zeta$ Pup} \citep{ReH96},  \object{$\xi$~Per} and \object{$\lambda$ Cep} \citep{dJHS99}, and  \object{HD 152249} \citep{GSL08}. Additionally, \citet{FGB91} suggested that the O7~II star \object{HD 34656} shows radial pulsations, possibly revealing the radial fundamental mode.


 \object{HD 152219} is an O9~III + B1-2~III/V short period binary ($P\sim4.2$~d) in the core of the NGC 6231 cluster. In \citet[][ Paper~I]{SGR06_219}, we showed that the primary component displays clear line profile variability (LPV). With $\log T_\mathrm{eff}=4.50$ and $\log L/L_\odot=5.07$, the primary star lies on the prolongation of the $\beta$~Cep instability strip \citep[e.g., ][]{Pam99, MMD07}, within the domain of instability predicted by \citet{Pam99} for high-luminosity stars. 

In the present paper, we report on a 4-night intensive monitoring campaign of the object, that aimed at determining the nature of the observed LPVs. In Paper~I, we focused on the \heb\l4686 line as it is the only well isolated, strong primary line that is uncontaminated by the secondary signature or by neighbouring lines. This strategy not only renders the analysis more straightforward, but it has the further advantage of avoiding any poorly quantified effects that can hamper spectral disentangling methods in the presence of varying spectral signatures \citep[see e.g., ][]{LRM08}. In the present paper, we analyze a larger subset of lines: \hbeta, \hea\ll4712, 4921, 5875, 6678 and \heb\l4686.

\begin{table}
\caption{Synthetic view of the June 2006 \fer\ campaign. The first column indicates the night number with respect to the beginning of the campaign. The second and third columns yield the heliocentric Julian date (in format HJD$-$2\,450\,000) and phase interval ranges covered. The last column lists the number $n$ of spectra obtained each night. }
\label{tab: opt_diary2}
\centering
\begin{tabular}{c c c c c }
\hline\hline
\vspace*{-3mm} \\
 \# & HJD range & $\phi_{\lambda4686}$ & $n$\\
\hline
 1 & 3911.50709--3911.86854 & 0.754--0.839 & 19 \\
 2 & 3912.47729--3912.87965 & 0.983--0.078 & 32 \\
 3 & 3913.47072--3913.88523 & 0.217--0.315 & 49 \\
 4 & 3914.48017--3914.87588 & 0.455--0.548 & 34 \\
\hline			        
\end{tabular}		           			 
\end{table} 

\begin{table}
\caption{Sample of the journal of the spectroscopic observations of \hda. `HJD' gives heliocentric Julian date (HJD$-$2\,450\,000) at mid-exposure. $\phi_{\lambda4686}$ indicates the corresponding phase according to the new \heb\l4686\ ephemeris of Table \ref{tab: orbit}. $ RV_1$  gives the primary radial velocity  as measured on the \heb\l4686 line. `Exp' lists the exposure time (in seconds) while `SNR' indicates the measured signal-to-noise ratio in the range 4740--4750\AA. The full table is available at the CDS.} 
\label{tab: opt_diary}\small
\centering
\begin{tabular}{rrrrr  }
\hline\hline
\vspace*{-3mm} \\
HJD & $\phi_{\lambda4686}$  & $ RV_1$ & Exp & SNR \\
\hline
3860.81531 & 0.804  &     20.15  &  900 & 320 \\  
3861.66960 & 0.005  & $-$133.85  &  900 & 323 \\  
3861.92118 & 0.064  & $-$149.33  &  900 & 361 \\  
3862.65506 & 0.237  &  $-$69.29  &  900 & 284 \\  
3862.90246 & 0.296  &  $-$29.43  &  900 & 356 \\  
3863.64302 & 0.470  &     74.92  &  900 & 294 \\  
3863.88654 & 0.528  &     90.92  &  900 & 380 \\  
3864.65014 & 0.708  &     67.76  &  720 & 310 \\  
3864.86446 & 0.758  &     42.87  &  600 & 334 \\  
3911.50709 & 0.758  &     38.58  & 1200 & 142 \\   
3911.52338 & 0.762  &     38.77  & 1500 & 276 \\
3911.55914 & 0.771  &     29.30  &  900 & 215 \\
3911.57020 & 0.773  &     28.47  &  900 & 275 \\
3911.58127 & 0.776  &     27.06  &  900 & 223 \\
3911.61757 & 0.784  &     17.72  & 1200 & 194 \\
3911.63210 & 0.788  &     15.42  & 1200 & 229 \\
3911.64664 & 0.791  &     15.23  & 1200 & 213 \\
3911.66996 & 0.797  &     14.08  & 1200 & 306 \\
3911.68451 & 0.800  &     13.18  & 1200 & 325 \\
3911.69905 & 0.804  &     14.59  & 1200 & 368 \\
3911.71393 & 0.807  &     14.65  & 1200 & 276 \\
3911.72847 & 0.811  &     14.59  & 1200 & 285 \\
3911.76994 & 0.820  &     14.52  & 1200 & 302 \\
3911.79196 & 0.826  &     12.80  & 1020 & 249 \\
 \hline			        
\end{tabular}		           			 
\end{table}

\section{Observations and data reduction }\label{sect: obs}

During four consecutive nights in June 2006, we intensively monitored \hda, using the \fer\ echelle spectrograph mounted on the ESO/MPG 2.2m telescope at La Silla (Chile). The detector was a 2k $\times$ 4k EEV CCD with a pixel size of 15$\mu$m $\times$ 15$\mu$m. The spectral resolving power of \fer\ is 48\,000. The atmospheric dispersion corrector (ADC) was used to maximize the throughput at high airmass (up to $z<3$), but it was also used at low airmass to preserve the homogeneity of the data set. The exposure time was varied between about 300 and 1200 sec in order to match the observing conditions (airmass, atmosphere transparency, seeing) but also to cover various time ranges. The  campaign yielded a total of 134 spectra. The resulting signal-to-noise ratios (SNR), as measured on the reduced 1D spectra in the continuum region 4740--4750\AA, are all above 200, but for 10 spectra. On average, the SNR is close to 300.  Table \ref{tab: opt_diary2} provides a  synthetic view of the campaign while  Table \ref{tab: opt_diary} presents the journal of the observations. The full table is available at the CDS and, beyond the sample data displayed here, the table further contains all the RV measurements performed on the various lines studied in this paper

As a complement to the June 2006 campaign, nine \fer\ spectra were obtained during five consecutive nights in May 2006 in order to improve the accuracy of the ephemeris. The typical exposure time was 900~sec and the obtained SNR about 300 as seen from Table \ref{tab: opt_diary}.

The data reduction was performed using a modified \fer\ pipeline working under the \mid\ environment. Beyond the  modifications already described in Paper~I, several new features were implemented. We used a 2D fit of the order position for improved stability. We also used the wavelength calibration frames obtained with the `new' ThArNe calibration lamp \citep[see e.g., ][]{FEROS-UMa}. The latter however heavily saturates in the red, even on the shortest exposures provided by the calibration plan\footnote{This drawback could be circumvented by using the `old' ThAr+Ne lamp. However, its flux in the blue is very limited and one would need to add a large series of frames, an option which is not supported by the calibration plan either. }.  At the time of our observations, the daily \fer\ wavelength calibrations consisted of 3 series of two exposures with increasing exposure times of 3s, 15s and 30s, a scheme that is repeated at the begining and at the end of the night.  To optimize the wavelength calibration, we computed a master calibration frame for each exposure time, by summing the frames with identical exposure times. We then extracted the ThArNe calibration spectra separately for each master frame. We finally used the detected lines from the 3s/15s/30s master frames to respectively calibrate the orders 30--39/10--29/1--9. The saturated lines or the lines outside the linearity range of the detector were rejected from the fit and the detection threshold was optimized for each order. Finally, a 2-iteration 3-sigma clipping method was used to discard the very few remaining discordant lines (mostly because of artifacts related to poorly corrected bad column effects). The obtained wavelength calibration residuals were all in the range 2.7-2.9~m\AA. 


\begin{table*}
\begin{center}
\caption{ \label{tab: orbit}Orbital solutions deduced from different lines (Col.~1). The usual notations for the orbital elements have been used. Quoted uncertainties are the 1-$\sigma$ error bars.}
\begin{tabular}{c c c c c c c}
\hline
Line       &  $P$              &   $e$              &$\omega$                   &  $K_1$          & $\gamma_1$      & r.m.s.\\
           &  (d)              &                    &(\degr)                    & (\kms)          & (\kms)          & (\kms)\\
\hline
\hbeta        &  4.24021 & 0.058 $\pm$ 0.006 & 140 $\pm$  5\hspace*{2mm}  & 117.0 $\pm$ 0.5 & $-29.6 \pm$ 0.6 &  4.9 \\

\heb\,\l4686  &  4.24029 & 0.085 $\pm$ 0.003 & 153 $\pm$  1\hspace*{2mm}  & 124.4 $\pm$ 0.2 & $-17.3 \pm$ 0.3 & \hspace*{1mm}2.0 \\

\hea\,\l4922  &  4.24023 & 0.057 $\pm$ 0.006 & 148 $\pm$  4\hspace*{2mm}  & 107.7 $\pm$ 0.4 & $-17.9 \pm$ 0.5 & \hspace*{1mm}3.8 \\

\hea\,\l5015  &  4.24015 & 0.059 $\pm$ 0.006 & 150 $\pm$  4\hspace*{2mm}  & 112.1 $\pm$ 0.4 & $-22.2 \pm$ 0.6 & \hspace*{1mm}4.0 \\
\hea\,\l5876  &  4.24021 & 0.069 $\pm$ 0.004 & 151 $\pm$  2\hspace*{2mm}  & 115.9 $\pm$ 0.3 & $-22.4 \pm$ 0.4 & \hspace*{1mm}2.9 \\
\hea\,\l6678  &  4.24022 & 0.061 $\pm$ 0.005 & 151 $\pm$  3\hspace*{2mm}  & 114.4 $\pm$ 0.4 & $-14.7 \pm$ 0.5 & \hspace*{1mm}3.2 \\
\hline 
\end{tabular}
\end{center}
\end{table*}

\begin{figure}
\centering
\includegraphics[width=\columnwidth]{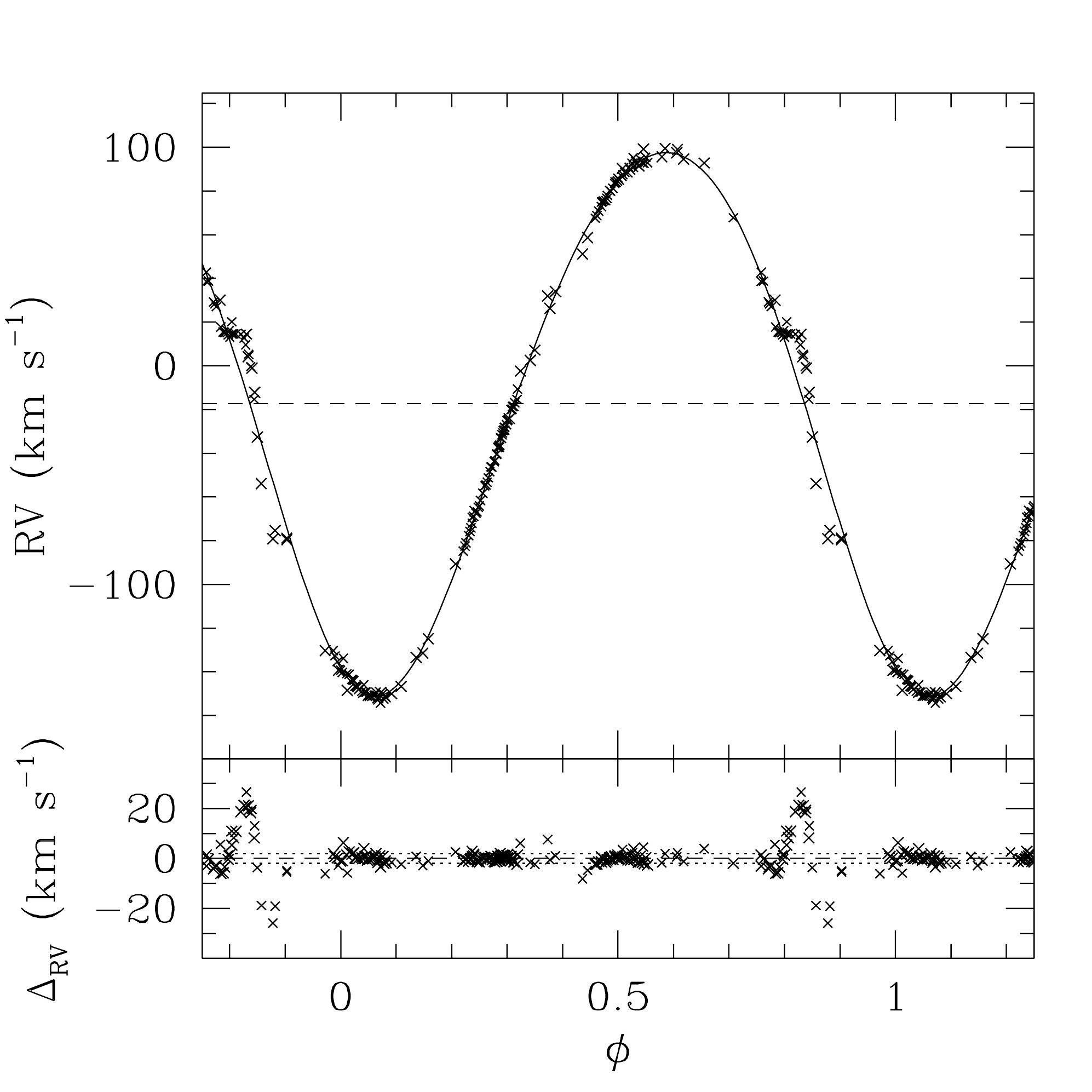}
\caption{{\it Upper panel:} \hda\ RV-curve  corresponding to the updated \heb\l4686 orbital solution of Table \ref{tab: orbit}. The dashed line indicates the systemic velocity. {\it Lower panel:} Residuals of the fit, in the sense data minus model. The dotted lines indicate the $\pm$ 1-$\sigma$ domain around the zero point of the solution (dashed line).  }
\label{fig: rv4686}
\end{figure}

\section{Updated orbital solutions }\label{ssect: orb_orb}

In Paper~I, we showed that radial velocity (RV) solutions derived from different spectral lines could show significant systematic differences. In particular the semi-amplitude of the  RV curves derived from different lines could differ by up to 10~\kms. As a necessary step in the LPV analysis (see Sect.~\ref{sect: lpv}) is to correct the observed spectra for the primary motion, we first computed updated orbital solutions for each of the considered spectral lines.

As for the data of Paper~I, we measured the Doppler shifts by fitting one or two Gaussians to the line profiles, depending whether the SB2 signature was visible or not. The RVs  were then computed adopting the same rest wavelengths as in \citetalias{SGR06_219} \citep{CLL77, Und94}. While other authors have showed that more accurate RVs can be obtained by using a cross-correlation technique \citep[e.g., ][]{RNC01}, the more straightforward Gaussian-fitting method still allows us to reach a RV accuracy of a few \kms\ or better, which is sufficient for our purpose. Beyond the present data set, we also used the data of  \citetalias{SGR06_219} to obtained the RVs of several lines that were not studied in \papI\ but that are analyzed here.

As our aim is to correct for the primary orbital motion, we exclusively focused on the primary orbital solution. Combining the current data with the data from \papI, we computed  new orbital solutions for each line, using  the Li\`ege Orbital Solution Package \citep[LOSP,][]{SG09}. The uncertainties quoted in Table~\ref{tab: orbit} were computed using the observed dispersion of a set of 1000 simulated solutions through the Monte-Carlo simulation module of the package.  The new orbital solutions are compatible, within the uncertainties, with those quoted in \papI. As discussed in Sect.~\ref{sect: lpv}, RV measurements performed in the phase interval $\phi_{\lambda4686}=0.71-0.95$ are not representative of the motion of the star mass center and, indeed, those points show larger residuals. In a second step, we thus recomputed the orbital solutions excluding observations obtained during that phase range. This significantly improved the residuals of the fit. Table \ref{tab: orbit} provides the updated orbital solutions for the different lines. The main noticeable difference with Paper~I is a reduced eccentricity. While the new values remain marginally within the 3-$\sigma$ errors bars of \papI, the observed differences can also be due to a slow apsidal motion as suggested by \citet{MHN08}. 

As the \heb\l4686 line is the only line in our sample that is uncontaminated by the secondary signature, and as the \heb\l4686 solution shows the lowest residuals, we will adopt the \heb\l4686 ephemeris in the rest of this paper. In the following, the periastron passage of the \heb\l4686 solution, at $T_0=2\,453\,997.338$ (HJD), is adopted as phase zero: $\phi_{\lambda4686}=0.0$.  The \heb\l4686 RV-curve is displayed in Fig.~\ref{fig: rv4686}. 

\begin{figure}
\centering
\includegraphics[width=\columnwidth]{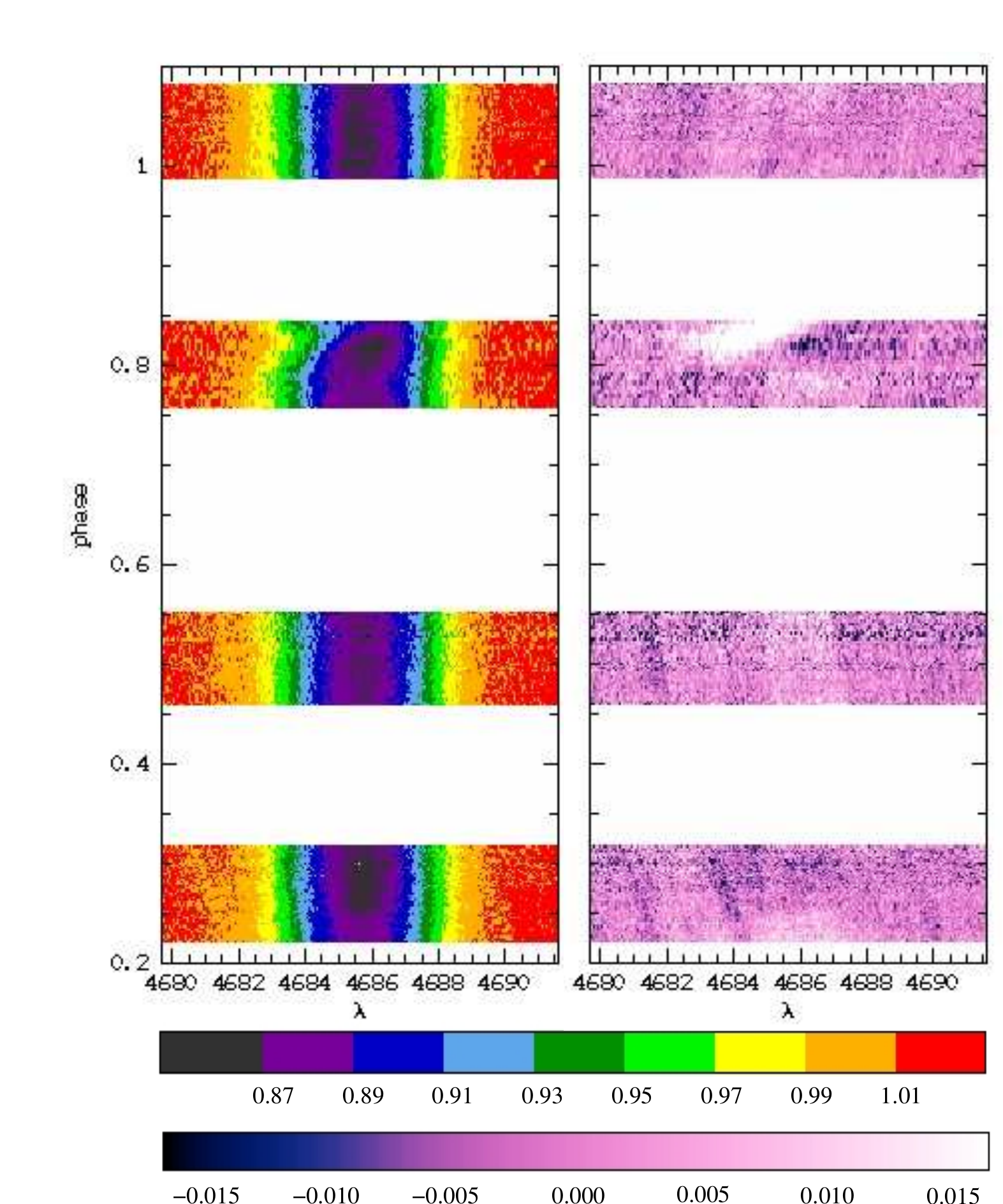}
\caption{{\it Left panel:} Color-coded image of the evolution of the \heb\l4686 line profile with the phase. Spectra are  displayed in the reference frame of the primary. {\it Right panel:} same as left panel for the difference spectra, i.e. the spectra {\it minus} the median spectrum computed over all non-eclipse phases. Color scales for both panels are indicated at the bottom of the image. }
\label{fig: diff2Dsolo}
\end{figure}

\begin{figure*}
\centering
\includegraphics[width=8cm,angle=-90]{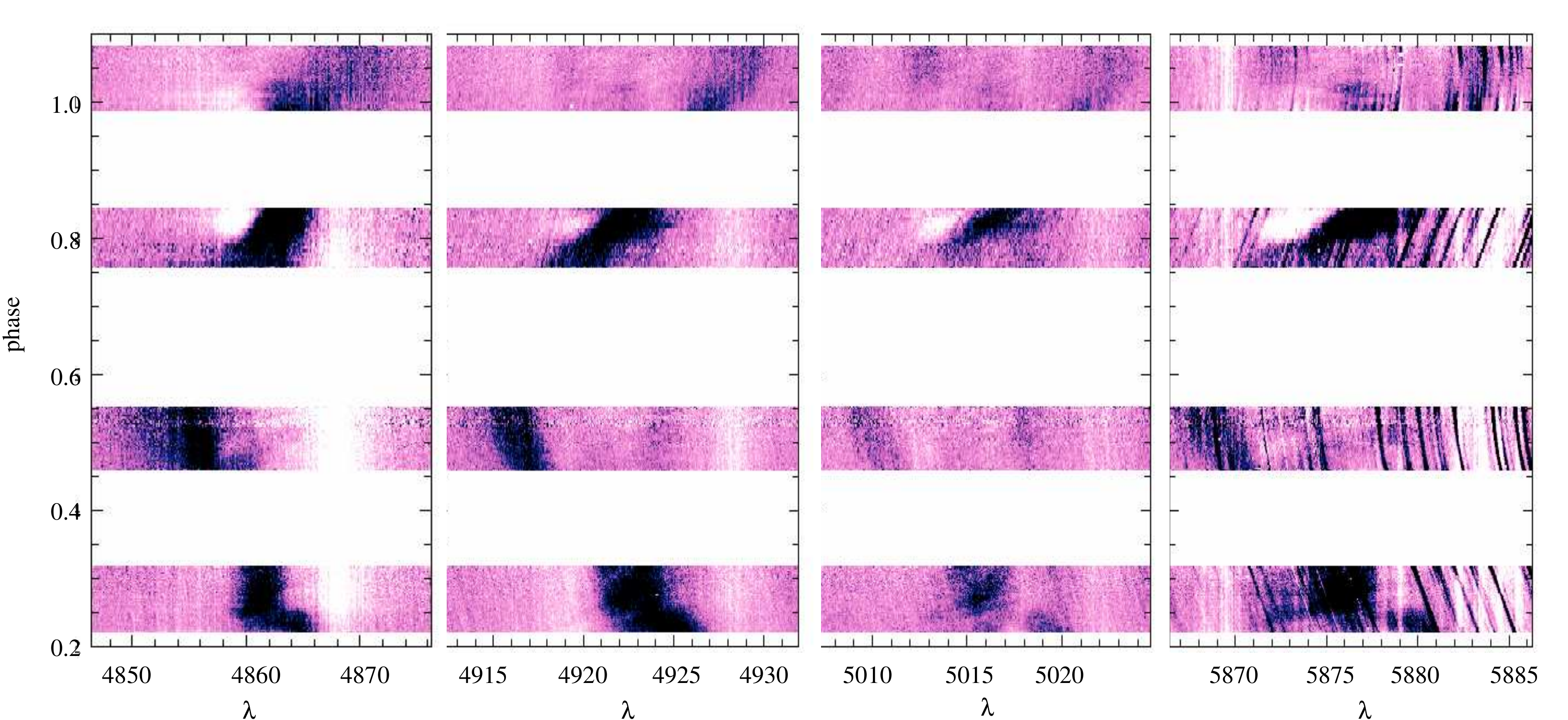}
\caption{ Evolution of the difference spectra for the \hbeta\ and \hea\l 4922, 5015, 5876 lines. The color scale is identical to the one of Fig.~\ref{fig: diff2Dsolo} (right panel).}
\label{fig: diff2D_all}
\end{figure*}

\section{Line profile analysis}\label{sect: lpv}

Using the updated orbital solutions, we corrected the spectra for the primary motion. As an example, Fig.~\ref{fig: diff2Dsolo} (left panel) shows the evolution of the \heb\l4686 line profile in the primary reference frame. It clearly illustrates that the main LPV is observed at $\phi_{\lambda4686}\sim0.79-0.91$, which corresponds to the time of the primary eclipse, when the companion is passing in front of the primary (an effect known as the Rossiter-McLaughlin effect). Additionally, one can also see a change of the intensity of the \heb\l4686 line during the secondary eclipse ($\phi_{\lambda4686}\sim0.23-0.37$), resulting from the changing dilution of the line by the varying continuum. However, there seems to be little or no profile  variations outside those phase intervals. 

To increase the dynamics of the image, we computed the median spectrum outside eclipse phases. The median spectrum was subsequently subtracted from all our spectra, yielding `difference spectra'. Fig.~\ref{fig: diff2Dsolo} (right panel) shows the evolution of \heb\l4686  difference spectra with phase. Clearly, no further LPV pattern can been seen at the $\approx0.005$ fraction of the continuum level.

Similarly, Fig.~\ref{fig: diff2D_all} shows the evolution, along with the phase, of the \hbeta\ and \hea\ll4922, 5015, 5876 difference spectra. The primary profile during nights \#1 and \#3 ($\phi_{\lambda4686}\sim0.8$ and 0.25) are clearly blended with the secondary signature. During nights \#2 and \#4 ($\phi_{\lambda4686}\sim1.05$ and 0.5), the separation between the two components is large enough, so that the primary profile can be searched for NRP signatures. As for the \heb\l4686 line however,  no profile  variations can be detected beyond the eclipse effects.

\begin{figure*}
\centering
\includegraphics[width=.6\columnwidth]{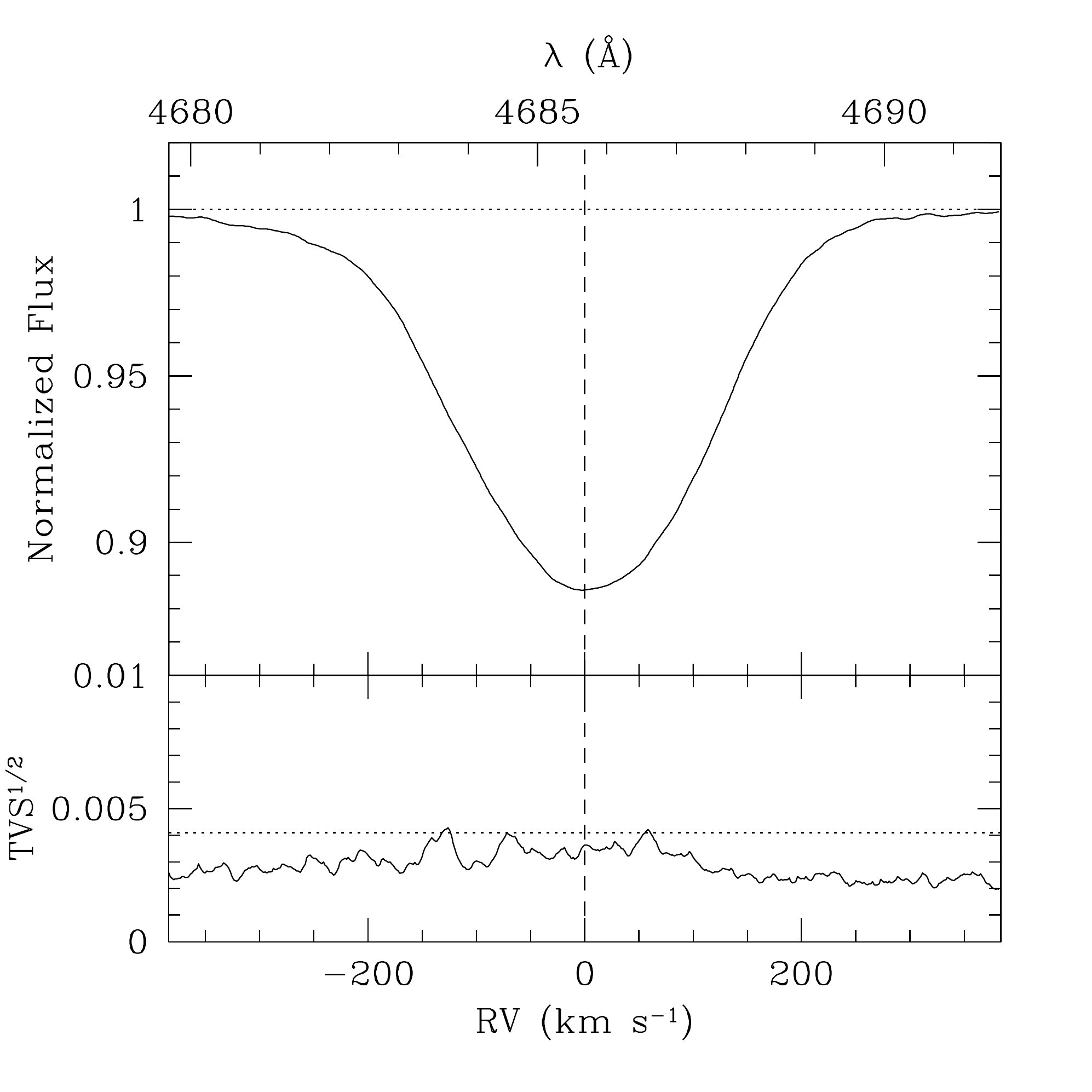}
\includegraphics[width=.6\columnwidth]{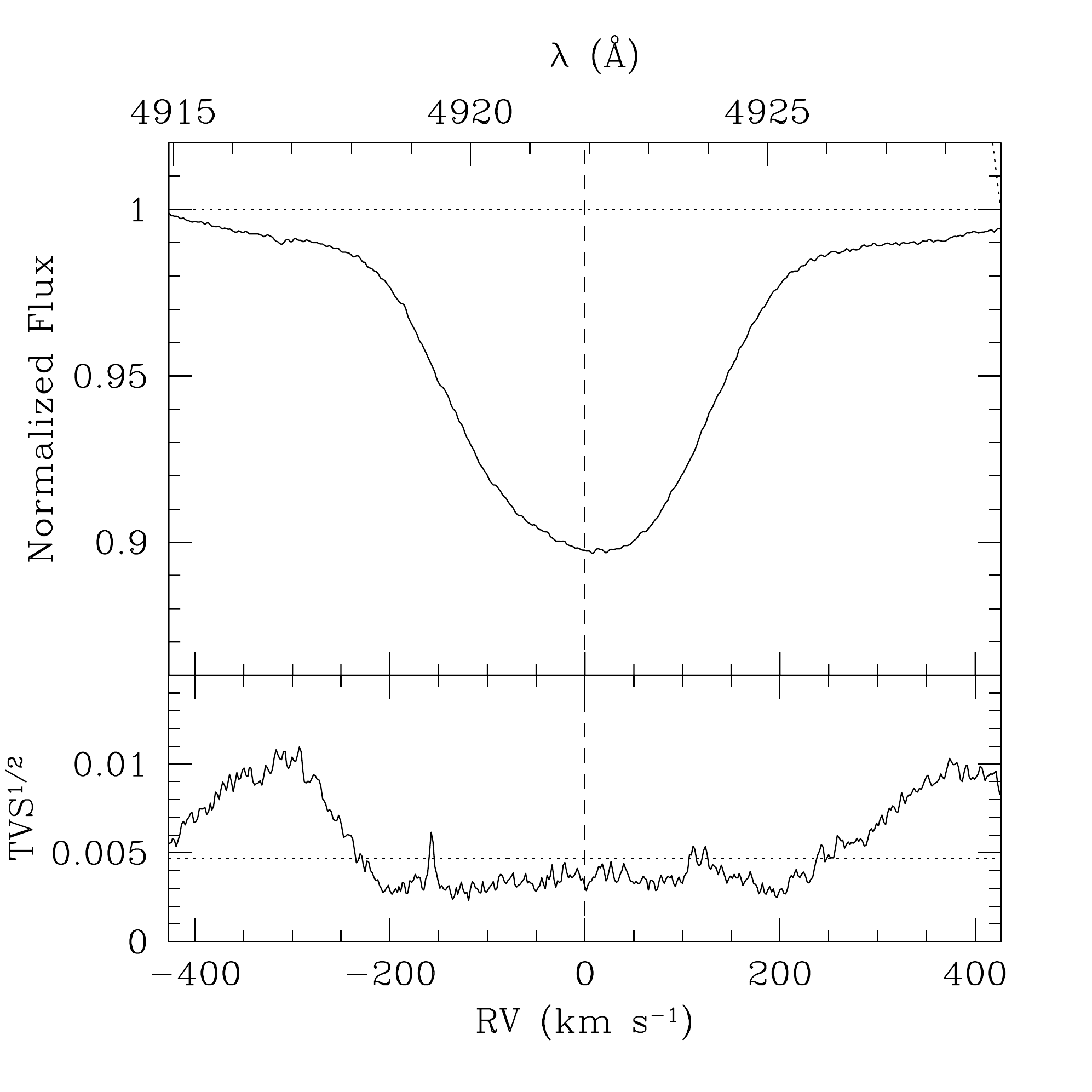}
\includegraphics[width=.6\columnwidth]{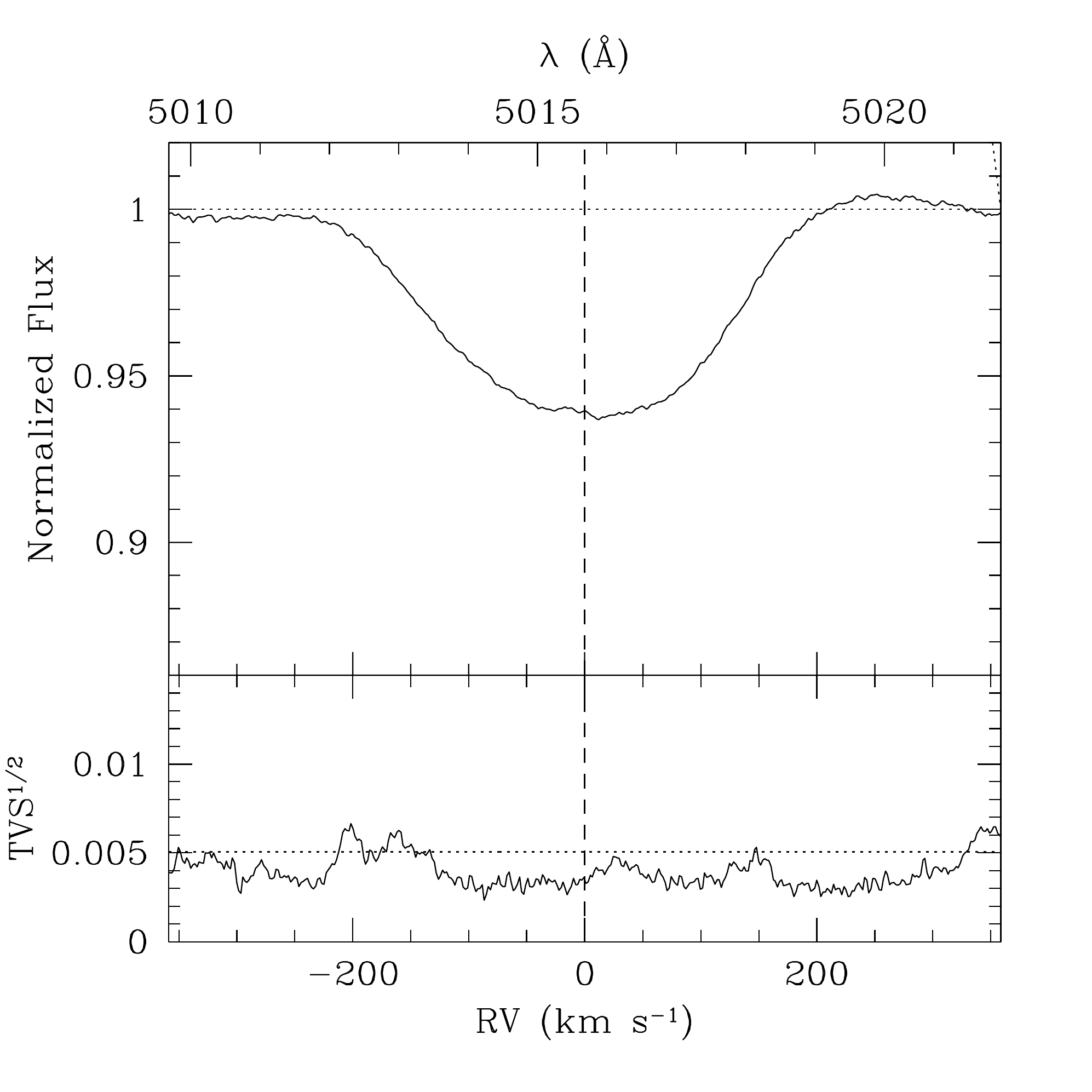}

\caption{Averaged line profile (upper panels) and square root of the TVS (lower panels) for three different lines. On the upper panels, the dotted lines show the continuum level while the dashed lines indicate the primary velocity frame. On the lower panels, the dotted lines indicate the variability threshold corresponding to a significance level of 0.001. }
\label{fig: TVS}
\end{figure*}

To quantitatively verify that no significant LPV are present, we computed the Time Variance Spectrum \citep[TVS, ][]{FGB96} using only spectra obtained during night \#2 and \#4, thus outside the eclipse phases. While we followed the original idea of \citet{FGB96}, we have weighted each pixel in the TVS by its standard deviation.  Fig.~\ref{fig: TVS} shows the averaged (outside eclipse) profiles of the  \heb\,\l4686 and \hea\ll4922, 5015 lines and the corresponding TVS spectra. Using the formalism proposed by \citet{FGB96}, we tested our data against the null hypothesis `not variable', adopting a significance level of 0.001. Beside some variations in the wings of the \hea\ lines due to the displacement of the secondary lines in the primary reference frame, it is clear from Fig.~\ref{fig: TVS} that the null hypothesis cannot be rejected. As a consequence, we consider the null hypothesis to be very likely and we conclude that no significant variability is seen in the primary line profile (except for the eclipse effects). 

We alse  performed a similar TVS analysis on the \hbeta\ and \hea\ll5876, 6678 lines, but these lines turned out to be inadequate for our purposes. The \hea\ll5876 line profile is polluted by atmospheric H$_2$O lines that  varies due to the different amount of water vapor along the observing run and to the large airmass spaned by the observations.  The \hea\ll6678 blend is located in a telluric-free window of the red spectrum. Unfortunately it is partly affected by residual fringing that is not fully corrected by the flat-fielding. Finally, the  \hbeta\ secondary line is so broad that, even at maximum separation, it still strongly contaminates the primary line profile. Despite the fact that the TVS is not applicable,  inspection of Fig.~\ref{fig: diff2D_all} still reveals  that no clear NRP signature can be seen in those lines.

While we now concludes that NRPs remain undetected in our observations, the high quality of the current data set allows us to put strong constraints on the upper limit of potential low-amplitude non-radial pulsations (NRPs), by using the square-root of the time variance. From Fig.~\ref{fig: TVS}, the noise level across the line profile is about 4$\times10^{-3}$ of the continuum level. As shown in  Paper I, the primary companion is expected to contribute to about 91\% of the continnum level. Assuming that the secondary is not contributing at all to the noise, we can set up the detection limit at a fraction of 0.005 of the primary continuum.


\begin{table}
\centering
\caption{Physical parameters of the known and suspected non-radial pulsators. }
\label{tab: pulsator}
\begin{tabular}{cccccc}
\hline\hline
Object & Sp.Type & $\log\frac{L}{L_\odot}$ & $\log T_\mathrm{eff}$ & $\log g$  & Status$^a$\\
       &         &                         & (K)                    & (cm\,s$^{-2})$\\
\hline
HD~93521      & O9.5V   & 4.62 &  4.48 & 3.92    & C\\
$\zeta$~Oph   & O9V     & 4.72 &  4.50 & 3.92    & C\\
HD~152219$^b$ & O9.5III & 5.21 &  4.50 & 3.51    & N\\
$\xi$~Per     & O7.5III & 5.36 &  4.52 & 3.59    & C\\
HD~152249     & O9I     & 5.54 &  4.47 & 3.23    & S\\
$\lambda$~Cep & O6I     & 5.78 &  4.55 & 3.48    & C\\
$\zeta$~Pup   & O5I     & 5.87 &  4.59 & 3.57    & S\\
\hline
\end{tabular}\\
\flushleft
$^a$ C: confirmed NRPs, S: suspected NRPs, N: no NRP \\
$^b$ Only the primary component is considered here.\\
\end{table}

\section{Discussion} \label{sect: ccl}

The analysis of a high-quality times series of the profiles of \hbeta, of \heb\l4686 and of various \hea\ lines in the \hda\ spectrum failed to reveal any significant LPV beside the Rossiter-McLaughlin effect. The data  displays a SNR close to 300 on average and have allowed us to  place an  upper limit on the variability level (and thus on the amplitude of possible NRPs in \hda) as low as a few parts per thousand of the continuum level.

 Table \ref{tab: pulsator} lists some of the known and suspected non-radial pulsators among O-type stars. The quoted physical properties are taken from \citet{MSH05}, based on the spectral type of the objects. Along the luminosity axis, \hda\ is roughly located between $\zeta$~Oph on one side and $\xi$~Per and HD~152249 on the other, in a yet poorly mapped region of the Hertzsprung-Russell diagram (HRD). \hda\ thus lies  roughly in-between the confirmed main sequence non-radial pulsators and the suspected brighter giant and supergiant ones. Yet, this luminosity sequence does not correlate well with the observations of NRPs. This might thus suggest the existence of two different regimes for the instability domain for massive stars, separated by a region where the NRPs are inhibited or have a much smaller amplitude. Results from on-going theoretical work are eagerly awaited, to revise the expected instability domain of the high-mass stars and to identify the key parameters that govern the pulsational processes in massive stars.

Because very few high quality monitorings of massive stars exist so far, the upper limit derived in the present paper can further be used to test asteroseismology models and to help to accurately constrain the exact location of the high-luminosity end of the instability strip in the Hertzsprung-Russell diagram.

\begin{acknowledgements}
The author is grateful to E. Gosset, H. Hensberge and G. Rauw for comments on the manuscript. We are also pleased to thank the referee, Dr. Bolton, whose comments help to improve the quality of the paper.
\end{acknowledgements}

\bibliographystyle{aa}
\bibliography{/home/hsana/LITERATURE/literature.bib}

\begin{thebibliography}{19}
\expandafter\ifx\csname natexlab\endcsname\relax\def\natexlab#1{#1}\fi

\bibitem[{{Conti} {et~al.}(1977){Conti}, {Leep}, \& {Lorre}}]{CLL77}
{Conti}, P.~S., {Leep}, E.~M., \& {Lorre}, J.~J. 1977, \apj, 214, 759

\bibitem[{{de Jong} {et~al.}(1999){de Jong}, {Henrichs}, {Schrijvers}, {Gies},
  {Telting}, {Kaper}, \& {Zwarthoed}}]{dJHS99}
{de Jong}, J.~A., {Henrichs}, H.~F., {Schrijvers}, C., {et~al.} 1999, \aap,
  345, 172

\bibitem[{{Fullerton} {et~al.}(1991){Fullerton}, {Gies}, \& {Bolton}}]{FGB91}
{Fullerton}, A.~W., {Gies}, D.~R., \& {Bolton}, C.~T. 1991, \apjl, 368, L35

\bibitem[{{Fullerton} {et~al.}(1996){Fullerton}, {Gies}, \& {Bolton}}]{FGB96}
{Fullerton}, A.~W., {Gies}, D.~R., \& {Bolton}, C.~T. 1996, \apjs, 103, 475

\bibitem[{{Gosset} {et~al.}(2008){Gosset}, {Sana}, {Linder}, \& {Rauw}}]{GSL08}
{Gosset}, E., {Sana}, H., {Linder}, N., \& {Rauw}, G. 2008, in Comm. In
  Asteroseismolgy: Contribution to the Proceedings of the 38th LIAC, Vol. in
  press

\bibitem[{{Howarth} {et~al.}(1998){Howarth}, {Townsend}, {Clayton},
  {Fullerton}, {Gies}, {Massa}, {Prinja}, \& {Reid}}]{HTC98}
{Howarth}, I.~D., {Townsend}, R.~H.~D., {Clayton}, M.~J., {et~al.} 1998,
  \mnras, 296, 949

\bibitem[{{Linder} {et~al.}(2008){Linder}, {Rauw}, {Martins}, {Sana}, {De
  Becker}, \& {Gosset}}]{LRM08}
{Linder}, N., {Rauw}, G., {Martins}, F., {et~al.} 2008, \aap, 489, 713

\bibitem[{{Martins} {et~al.}(2005){Martins}, {Schaerer}, \& {Hillier}}]{MSH05}
{Martins}, F., {Schaerer}, D., \& {Hillier}, D. 2005, \aap, 436, 1049

\bibitem[{{Mayer} {et~al.}(2008){Mayer}, {Harmanec}, {Nesslinger}, {Lorenz},
  {Drechsel}, {Morrell}, \& {Wolf}}]{MHN08}
{Mayer}, P., {Harmanec}, P., {Nesslinger}, S., {et~al.} 2008, \aap, 481, 183

\bibitem[{{Miglio} {et~al.}(2007){Miglio}, {Montalb{\'a}n}, \&
  {Dupret}}]{MMD07}
{Miglio}, A., {Montalb{\'a}n}, J., \& {Dupret}, M.-A. 2007, Communications in
  Asteroseismology, 151, 48

\bibitem[{{Pamyatnykh}(1999)}]{Pam99}
{Pamyatnykh}, A.~A. 1999, Acta Astronomica, 49, 119

\bibitem[{{Pritchard}(2005)}]{FEROS-UMa}
{Pritchard}, J. 2005, FEROS-II User Manual v. 77.0 (LSO-MAN-ESO-22200-0001),
  European Southern Observatory

\bibitem[{{Rauw} {et~al.}(2008){Rauw}, {De Becker}, {van Winckel}, {Aerts},
  {Eenens}, {Lefever}, {Vandenbussche}, {Linder}, {Naz{\'e}}, \&
  {Gosset}}]{RDBvW08}
{Rauw}, G., {De Becker}, M., {van Winckel}, H., {et~al.} 2008, \aap, 487, 659

\bibitem[{{Rauw} {et~al.}(2001){Rauw}, {Naz{\'e}}, {Carrier}, {Burki},
  {Gosset}, \& {Vreux}}]{RNC01}
{Rauw}, G., {Naz{\'e}}, Y., {Carrier}, F., {et~al.} 2001, \aap, 368, 212

\bibitem[{{Reid} \& {Howarth}(1996)}]{ReH96}
{Reid}, A.~H.~N. \& {Howarth}, I.~D. 1996, \aap, 311, 616

\bibitem[{{Sana} \& {Gosset}(2009){Sana} \& {Gosset}}]{SG09}
{Sana}, H., \& {Gosset}, E. 2009, \aap, submitted

\bibitem[{{Sana} {et~al.}(2006){Sana}, {Gosset}, \&
  {Rauw}}]{SGR06_219}
{Sana}, H., {Gosset}, E., \& {Rauw}, G. 2006, \mnras, 371, 67

\bibitem[{{Underhill}(1994)}]{Und94}
{Underhill}, A.~B. 1994, \apj, 420, 869

\bibitem[{{Walker} {et~al.}(2005){Walker}, {Kuschnig}, {Matthews}, {Reegen},
  {Kallinger}, {Kambe}, {Saio}, {Harmanec}, {Guenther}, {Moffat}, {Rucinski},
  {Sasselov}, {Weiss}, {Bohlender}, {Bo{\v z}i{\'c}}, {Hashimoto},
  {Koubsk{\'y}}, {Mann}, {Ru{\v z}djak}, {{\v S}koda}, {{\v S}lechta}, {Sudar},
  {Wolf}, \& {Yang}}]{WKM05}
{Walker}, G.~A.~H., {Kuschnig}, R., {Matthews}, J.~M., {et~al.} 2005, \apjl,
  623, L145

\end{thebibliography}

\end{document}